\documentclass[aps,twocolumn,10pt,aps,amsmath,amssymb,nofootinbib,superscriptaddress,showkeys,showpacs]{revtex4-1}
\usepackage{epsfig} 
\usepackage{amsmath} 
\usepackage{graphicx}
\usepackage{color}
\usepackage{float}
\usepackage{soul,xcolor}
\usepackage[font=scriptsize]{caption}
\usepackage{braket}
\usepackage{bbold}
\usepackage{subfig}
\usepackage{braket}
\usepackage{titlesec}
\usepackage{placeins}
\usepackage{caption}
\begin{document} 

\title{Quantum correlations in neutrino oscillations in curved spacetime}

\author{Khushboo Dixit}
\email{dixit.1@iitj.ac.in}
\affiliation{Indian Institute of Technology Jodhpur, Jodhpur 342011, India}

\author{Javid Naikoo}
\email{naikoo.1@iitj.ac.in}
\affiliation{Indian Institute of Technology Jodhpur, Jodhpur 342011, India}

\author{Banibrata Mukhopadhyay}
\email{bm@iisc.ac.in}
\affiliation{Indian Institute of Science, Bangalore 560012, India}

\author{Subhashish Banerjee}
\email{subhashish@iitj.ac.in}
\affiliation{Indian Institute of Technology Jodhpur, Jodhpur 342011, India}

\date{\today}

\begin{abstract}
 Gravity induced neutrino-antineutrino oscillations are studied in the context of one and two flavor scenarios. This allows one to investigate the particle-antiparticle correlations in two and four level systems, respectively. Flavor entropy is used to probe the entanglement in the system.  The well known witnesses of non-classicality such as Mermin and Svetlichly inequalities are investigated. Since the extent of neutrino-antineutrino oscillation is governed by the strength of the gravitational field, the behavior of non-classicality shows interesting features as one varies the strength of the gravitational field. Specifically, the suppression of the entanglement with the increase of the gravitational field is observed which is witnessed in the form of decrease in the flavor entropy of the system. The features of the Mermin and the Svetlichny inequalities allow one to make statements about the degeneracy of neutrino mass eigenstates.
\end{abstract}

\maketitle

\section{Introduction}
The phenomenon of neutrino oscillation is well known and establishes the  non-zero mass of neutrinos. This idea was first introduced by Pontecorvo to explain the  solar neutrino problem and later  experimentally confirmed by Super-Kamiokande \cite{SuperK} and Sudbury Neutrino Observatory (SNO) \cite{SNO}. Till date, several experiments using solar, atmospheric, reactor and accelerator neutrinos have analyzed the data to calculate the oscillation parameters such as mixing angles and mass squared differences. Some of the upcoming experiments are planning to resolve some queries in neutrino sector like, CP-violating phase \cite{Schechter1981}, type of mass hierarchy, absolute mass scale of neutrinos \cite{kobzarev1981sum} and existence of sterile neutrinos. 
	
The effect of gravitational field on neutrino oscillation was also studied \cite{bm2}. It was shown that while a nonzero (Majorana) mass of neutrinos is required for the neutrino-antineutrino mixing, for oscillation between neutrino and antineutrino (or their mass eigenstates) to occur, their energies must be split. This splitting of energy due to gravitational effect, ``gravitational Zeeman effect", also gives rise to the effective charge-parity-time reversal  violation, as the effective masses (pertaining to the mass eigenstates)  are different \cite{bm1}. Hence, along with the lepton number violation in the neutrino sector (due to Majorana mass), gravitational Zeeman effect leads to neutrino and antineutrino oscillation. In fact, gravitational Zeeman effect has many other consequences, e.g. neutrino asymmetry, baryogenesis, experimental tests of curvature couplings of spinors etc. \cite{bmprd02,bmmpla05,bmujjal06,mavromatosepjc13,mosqueraapj17}. Moreover, the geometric phase in neutrinos has been a part of various studies in this context \cite{BLASONE1999,He2005,Wang2001,Dajka2011,Mehta2009,Dixit_2018}. It is well-known that the neutrinos propagating in a varying magnetic field acquire a geometric phase \cite{joshi2016}. However, the Zeeman like splitting was also shown to give rise to a geometric phase where the space-time curvature plays the role of the magnetic field \cite{bm2018}.

Gravitational field also modifies the mixing. In absence of gravity, this mixing is \textit{passive} in the sense that the value of mixing angle is always $\pi/4$. The effect of gravity leads to \textit{active} mixing, which depends on the strength of charge-parity-time reversal violation being determined by the strength of gravity itself. The gravity induced neutrino-antineutrino mixing also affects the flavor oscillations, thereby leading to modified neutrino oscillation even in the flavor sector. This naturally invites  the investigation of different aspects of neutrino oscillations. In this direction, the present work is devoted to analyzing the neutrino-antineutrino mixing using various tools of quantum foundations.

Among the celebrated notions of quantum foundations are  \textit{realism} and  \textit{locality}. The former says that the existence of an observable quantity does not depend on the observer, while the latter holds that  the information cannot  reach \textit{instantaneously} from one point of space to another.  These two concepts were used by J. Bell to develop the famous Bell inequity (BI) \cite{Bell}, the violation of which rules out the hypothesis of the hidden variable theory as an alternative of quantum mechanics.  The Bell type inequalities have been a subject matter of various works, for example, in optical and electronic systems \cite{Aspect,Tittel,Lanyon,Chakrabarty}, in particle physics systems such as mesons \cite{MacKenzie,Naikoo1} and neutrinos \cite{Caban,Bramon,Alok,Banerjee,Cervera,Kerbikov,Fu,Richter,Dixit}. A time analog of the  Bell inequality, known as Legget-Garg inequality has  gained a lot of attention recently \cite{Nikitin,Naikoo:2017fos}. The various avatars of Bell inequality have been developed  to  the study of nonlocality in a multipartite (more than two) systems, such as Mermin and Svetchlichny inequalities. Such manifestation of BI becomes important, for example, when dealing with the three flavors neutrino oscillation. \par

In this work, we explore various spatial quantum correlations in neutrinos propagating and oscillating in curved spacetimes. We have considered one and two-flavor neutrino cases, which lead to two- and four-level systems of neutrino-antineutrino oscillation, respectively. To study the correlation measures, the idea of mode entanglement is used \cite{blasone}, as discussed ahead.

The plan of this paper is as follows. In section II we briefly review the gravitational Zeeman effect. This is followed by a discussion of quantum correlations in one and two-flavor neutrino scenarios in section III. Section IV is devoted to the summary and conclusion.

\begin{figure}[t]
	\begin{minipage}{.9\linewidth}
		\centering
		\label{main:a}\includegraphics[width=70mm]{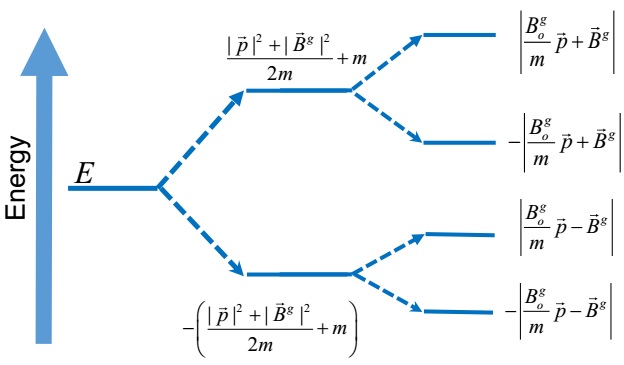}
	\end{minipage}%
	\vspace{0.10cm}
	\caption{Gravitational ``Zeeman-splitting": Weak gravitational effect, as given by Eq. (\ref{dirge2}), is considered for the ease of demonstration.}
	\label{grave}
\end{figure}

\section{Gravitational ``Zeeman effect"}
Dirac equation in the presence of background gravitational fields, in a local inertial coordinate, reduces as (see, e.g., \cite{birrell,kaku,bm1,schw})
\begin{eqnarray}
\left[i\gamma^\mu\partial_\mu-m+i\gamma^\mu A^g_\mu+\gamma^\mu \gamma^5 B^g_\mu\right]\psi=0,
\label{dirg}
\end{eqnarray}
where $A^g_\mu$ and $B^g_\mu$ are the gravitational 4-vector potentials (gravitational coupling with the spinor), $m$ is the mass of spinor, and $\gamma^5=\gamma_5=i\gamma^0\gamma^1\gamma^2\gamma^3$ as usual. Here we choose $\hbar=c=1$. For simplicity, in rest of the discussion we retain and explore the consequence of the axial-vector-like term only in equation which suffices for the present purpose. Nevertheless, the vector-like term might be anti-Hermitian as well 
in a local coordinate (depending on the spacetime nature) and can be removed from the total Lagrangian when added to  its complex conjugate part. 
This is particularly so for Majorana neutrinos,
when massive neutrinos are most plausibly believed to be Majorana typed.
Nevertheless, such a vector-like anti-Hermitian term would not contribute to 
the effective energy of the particle with an appropriate definition of 
dot-product in curved spacetime \cite{schw,parker2009}.

Now for the nontrivial solution of $\psi$, the Hamiltonians of the spin-up and spin-down particles are given by
\begin{eqnarray}
(H+\vec{\sigma}.\vec{B}^g)^2={\vec p}^2+{B_0^g}^2+m^2-2B_0^g{\vec \sigma}.{\vec p},
\label{dirge}
\end{eqnarray}
where $B^g_0$ is the temporal component of $B^g_\mu$. In the regime of weak gravity and when $m$ is much larger than the rest of the terms in the R.H.S. of Eq. (\ref{dirge}), it reduces to
\begin{eqnarray}
H=-\vec{\sigma}.\vec{B}^g\pm\left[\frac{{\vec p}^2+{B_0^g}^2}{2m}+m
-\frac{B_0^g{\vec \sigma}.{\vec p}}{m}\right].
\label{dirge2}
\end{eqnarray}

There are two-fold splits in dispersion energy, governed by two terms associated with the Pauli spin matrix, between up and down spinors for positive and negative energy solutions. See Fig. \ref{grave} demonstrating the same. 

In order to have nonzero $B^g_\mu$, spherical symmetry has to be broken, hence in Schwarzschild geometry it vanishes. 
 In Schwarzschild metric, any possible effect would arise from $A^g_\mu$, 
which is  removed in the present formalism. Indeed it is known 
\cite{0907.4367,0601095} that
spin evolution in spherical symmetric spacetime could arise only from an imaginary  Lorentz vector-like term.
On the other hand, in Kerr geometry $B^g_\mu$ survives. Also it survives in, e.g., early universe under gravity wave perturbation, Bianchi II, VIII and IX anisotropic universes. Note that in an expanding universe, gravitational potential $B^g_\mu$ turns to be constant at a given epoch which could act as a background effect.

In Kerr-Schild coordinate, after putting $\hbar$ and $c$ appropriately, the temporal
part of $B^g_\mu$ reads as 
\begin{equation}
B^g_0=-\frac{4az}{\bar{\rho}^2\sqrt{2r^3}}\frac{\hbar c}{r_g},
\end{equation}
where $\bar{\rho}^2=2r^2+a^2-x^2-y^2-z^2$, $r$ is the radial coordinate of the system expressed in units of $r_g$, $r_g=GM/c^2$, $M$ and $a$ (varying from $-1$ to $+1$) are respectively mass and dimensionless angular momentum per unit mass of the black hole, $G$, $c$ and $\hbar$ are respectively Newton's gravitation constant, speed of light and reduced Planck's constant. Naturally, $B^g_0$ survives (and is varying with space coordinates) for any spinning black hole leading to gravitational Zeeman effect.

In Bianchi II spacetime with, e.g., equal scale-factors in all directions, $B^g_0$ survives as 
\begin{equation}
B^g_0=\frac{4+3y^2-2y}{8+2y^2} \hbar c.
\end{equation}

\section{Quantum correlations in neutrinos}
\begin{figure}[h]
	\centering
	\includegraphics[width=50mm]{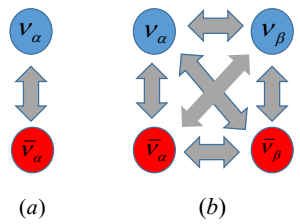}
	\caption{Neutrino-antineutrino oscillations in (a) one-flavor, and (b) two-flavor scenarios.}
	\label{NeuAnti}
\end{figure}

We are going to analyze the neutrino anti-neutrino oscillations in one and two flavor scenarios, which can be viewed as two and four level systems, respectively. The schematic diagram is given in Fig. \ref{NeuAnti}.

\begin{figure}
	\centering
	(a)\includegraphics[scale=0.46]{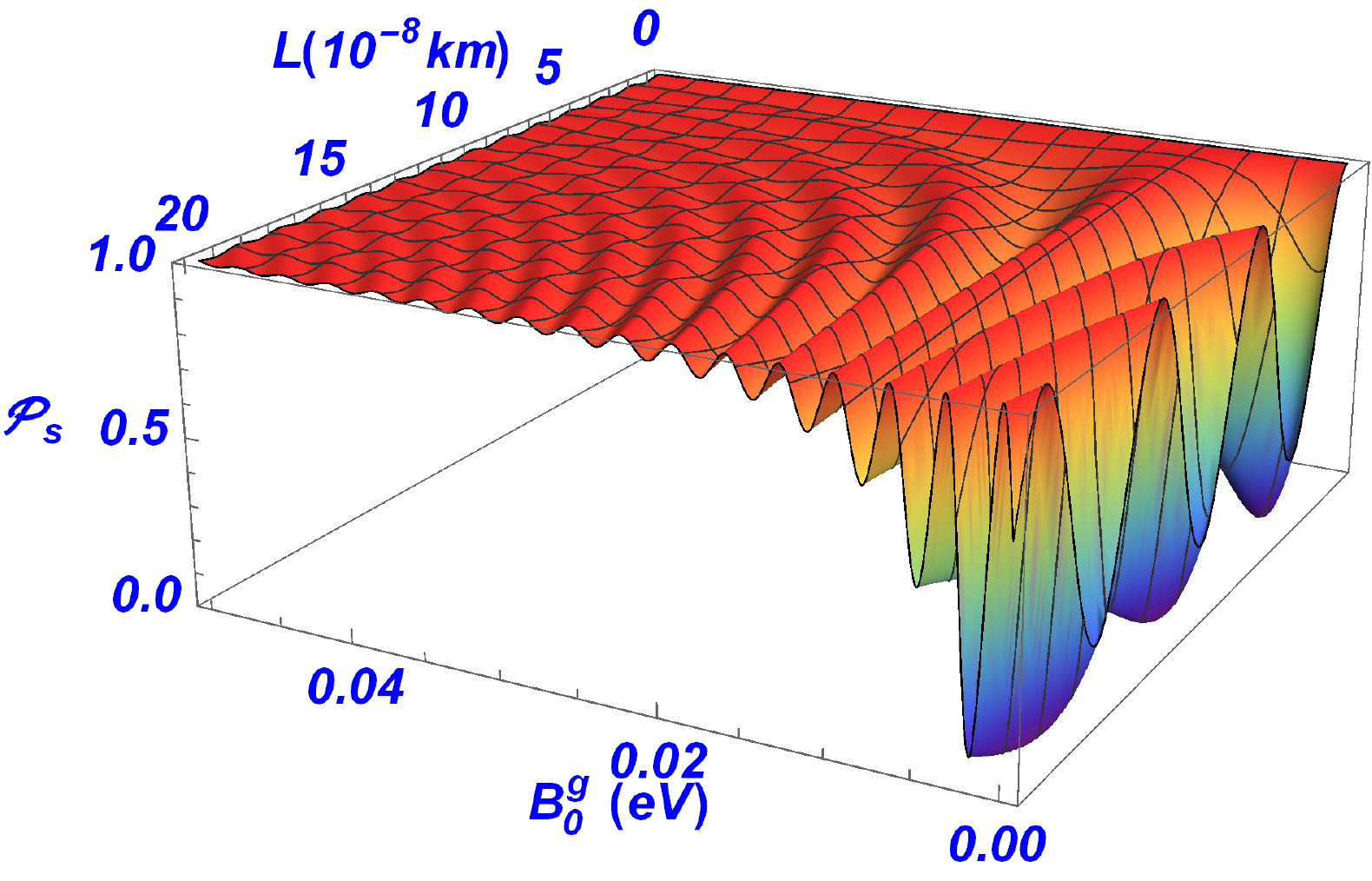} 
	(b)\includegraphics[scale=0.46]{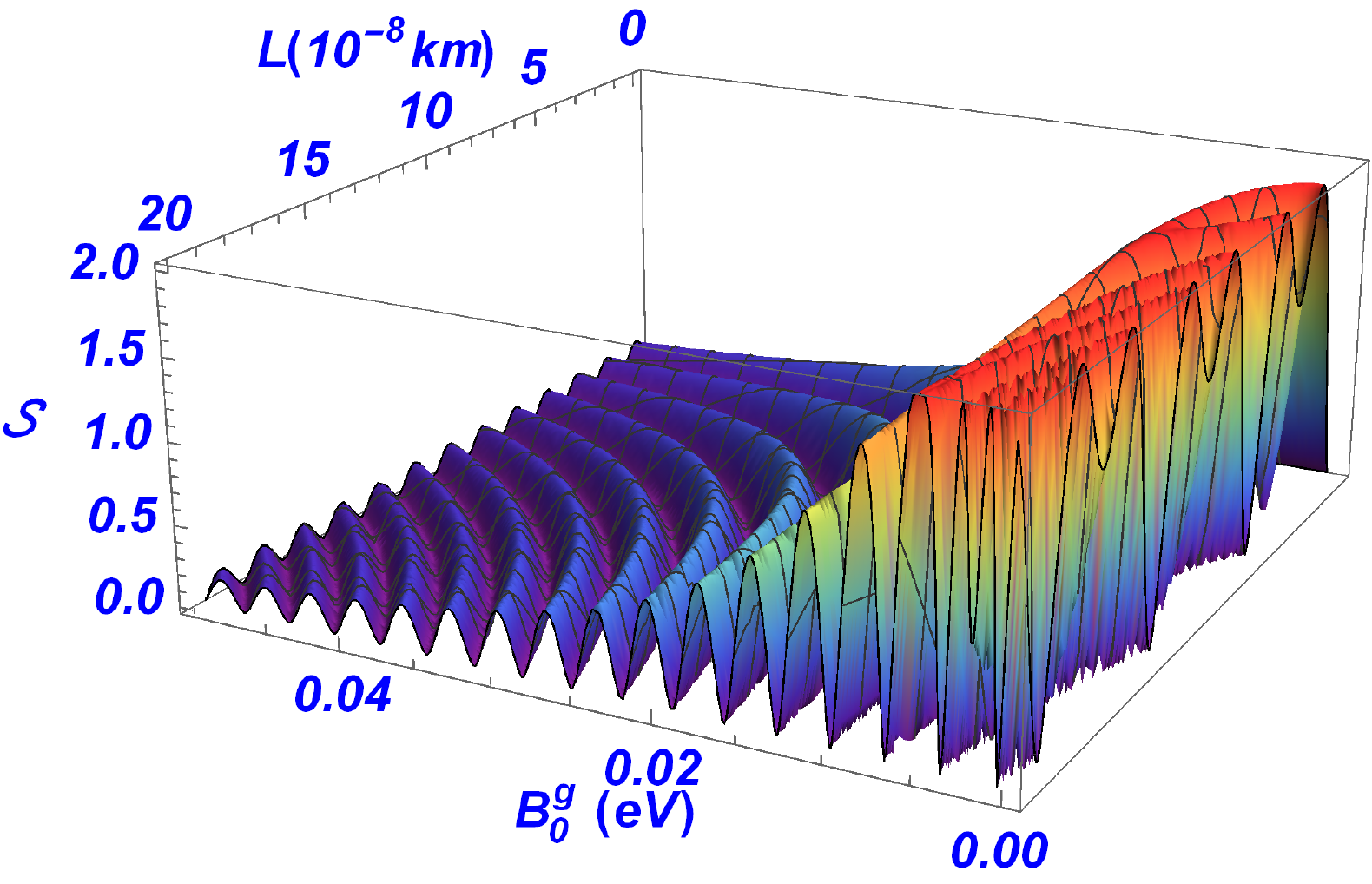}
	\caption{ Fixed flavor case: (a) Survival probability, and (b) von Neumann entropy, as functions of gravitational potential and distance traveled by neutrino/antineutrino, for the massive states $\nu_1 \leftrightarrow {\nu}_2$ corresponding to neutrino-antineutrino oscillation. The various parameters used are: $m = 5 \times 10^{-3}$ eV, $|\vec{B^g}|\sim 10^{-2}$ eV.}
	\label{Von}
\end{figure}
\FloatBarrier

\subsection{Neutrino-antineutrino mixing in single flavor scenario}
Let us consider, in Weyl representation, a 2-level system describing the mixing of neutrino ($\psi$) and antineutrino ($\psi^c$) \cite{baren} in the presence of gravitational coupling. We can express  the states
	with known mass (but unknown lepton number states, in the present case
	spin-states) in terms of states with known lepton/spin-states or vice versa. This is in the spirit of mixing in the neutral kaons, differing by two units of strangeness, whereas for neutrino and antineutrino it differs by two units of lepton number. The corresponding mass eigenstates for a particular flavor at $t=0$ are \cite{bm1,bm2}
\begin{eqnarray}
\label{mix}
|\nu_1(0)\rangle &=& \cos \theta~|\psi^c(0)\rangle~+~e^{i\phi}~\sin
\theta~|\psi(0)\rangle \nonumber \\
|\nu_2(0)\rangle &=& -\sin \theta~|\psi^c(0)\rangle~+~e^{i\phi}~\cos \theta~|\psi(0)\rangle, \end{eqnarray}
when \begin{equation} \tan\theta~=~\frac{m}{B_0^g+\sqrt{{B_0^g}^2+m^2}} \label{eqn20},\,\,\,\,\,\phi={\rm arg}(-m),
\end{equation} 
where $m$ is the Majorana mass of neutrino. Note that the mixing is maximum for $B_0^g = 0$. However, at an arbitrary time $t$ the mass eigenstates are
\begin{eqnarray}
\label{mixt}
|\nu_1(t)\rangle &=& \cos \theta~e^{-iE_{\psi^c} t}~|\psi^c(0)\rangle~+~e^{i\phi}~\sin
\theta~e^{-iE_\psi t}~|\psi(0)\rangle \nonumber \\
|\nu_2(t)\rangle &=& -\sin \theta~e^{-iE_{\psi^c} t}~|\psi^c(0)\rangle~
+~e^{i\phi}~\cos \theta~e^{-iE_\psi t}~|\psi(0)\rangle, \nonumber \\ 
\end{eqnarray}
where the dispersion energies for neutrino and antineutrino respectively, due to gravitational Zeeman-splitting, from Eq. (\ref{dirge}) are given by 
\begin{eqnarray}
\nonumber
E_\psi=\sqrt{(\vec{p}-\vec{B}^g)^2+m^2}+B_0^g,\\
E_{\psi^c}=\sqrt{(\vec{p}+\vec{B}^g)^2+m^2}-B_0^g.
\label{energy}
\end{eqnarray}

For ultra-relativistic neutrinos, $m \ll |\vec{p}|$ leading to $E_\psi - E_{\psi^c} \approx 2(B_0^g - |\vec{B}^g|)$, the survival probability of $\nu_1$ at time $t$ can be expressed as
\begin{equation}
\label{Prob}
\mathcal{P}_s(t) = 1 - \sin^2 2\theta \sin^2 \{(B_0^g - |\vec{B^g}|) (t)\}.
\end{equation}
Note that $(\nu_1(0), \nu_2(0))$ is just the transformed spinor of original $(\psi^c,\psi)$. In the limit of zero gravitational 
effect, i.e., $B_{\mu}^g \rightarrow 0$, $\mathcal{P}_s(t) \rightarrow 1$. Thus the neutrino-antineutrino oscillations primarily occur due to non-zero value of the gravitational potential, when the present analysis is performed for ultra-relativistic neutrinos.

In Fig. \ref{Von}a, the survival probability for $\nu_1 \leftrightarrow {\nu}_2$ oscillations is shown as a function of gravitational potential and the distance ($L \approx ct$ in ultra relativistic limit \footnote{Here, entire analysis is performed in local inertial coordinates. Hence, at each point, all the special relativistic norms are conveniently satisfied such that at a given local point, $B_{\mu}^g$ appears as constant background field \cite{bm2}.}) traveled by the neutrino/antineutrino. The survival probability can be seen to approach its maximum value unity as the gravitational potential increases. This implies that gravity suppresses the neutrino-antineutrino oscillations for a {\it  fixed $m$.} Further, as noted earlier, the neutrino-antineutrino oscillation approaches maximum when $B_0^g \rightarrow 0$.

\subsubsection{von Neumann entropy in oscillation}
The neutrino-antineutrino system can be treated as an effective two qubit system \cite{blasone, Alok, Banerjee, blasone2014} with the following \textit{occupation number} representation of states defined in Eq. (\ref{mix}) as
\begin{equation*}\label{occupation}
\ket{\nu_1(0)} \equiv \ket{10},  \qquad  \ket{\nu_2(0)} \equiv \ket{01}. 
\end{equation*}
The notation $\ket{10}$ amounts to asking whether we have a $\ket{\nu_1}$ state or not. In this notation, one can finally write 
\begin{eqnarray}\label{pure}
\ket{\nu_1(t)} &=& \mathcal{U}_{11}(t) \ket{10} + \mathcal{U}_{12}(t) \ket{01}, \nonumber \\
\ket{\nu_2(t)} &=& \mathcal{U}_{21}(t) \ket{10} + \mathcal{U}_{22}(t) \ket{01},
\end{eqnarray}
where the coefficients can be obtained from Eqs. (\ref{mix}) and (\ref{mixt}) as
\begin{align*}
\mathcal{U}_{11}(t) = \cos^2 \theta e^{-i E_{\psi^c} t} + \sin^2 \theta e^{-i E_{\psi} t}, \nonumber \\
\mathcal{U}_{12}(t) = \sin \theta \cos \theta (e^{-i E_{\psi} t} - e^{-i E_{\psi^c} t}), \nonumber \\
\mathcal{U}_{21}(t) = \sin \theta \cos \theta (e^{-i E_{\psi} t} - e^{-i E_{\psi^c} t}), \nonumber \\
\mathcal{U}_{22}(t) = \sin^2 \theta e^{-i E_{\psi^c} t} + \cos^2 \theta e^{-i E_{\psi} t}.
\end{align*}

A standard measure of entanglement for pure states, of form Eq. (\ref{pure}), is given by von Neumann entropy 

\begin{align}\label{entropy}
\mathcal{S} &= -\sum_{ \beta = 1, 2} |\mathcal{U}_{\alpha\beta}(t)|^2 \log_2 |\mathcal{U}_{\alpha\beta}(t)|^2 \nonumber \\& -\sum_{\beta = 1, 2} (1-|\mathcal{U}_{\alpha\beta}(t)|^2) \log_2 (1-|\mathcal{U}_{\alpha\beta}(t)|^2).
\end{align}  
Here $\alpha = 1~(2)$ corresponds to the $\nu_1$ ($\nu_2$) state. For the $\nu_1 \leftrightarrow \nu_2$ oscillation, Fig. \ref{Von}b shows the variation of $\mathcal{S}$ as a function of $B^g_0$ and $L$. The existence of entanglement is implied by $\mathcal{S} > 0$. The von-Neumann entropy, as an entanglement measure, is suitable for neutrino system since it can be expressed in terms of the survival and transition probabilities, which are experimentally measurable quantities \cite{Alok}. The increase in the gravitational potential is found to decrease the entanglement in the neutrino-antineutrino system for a {\it  fixed $m$.} Further, the entropy attains its maximum value when the survival (and
hence transition) probabilities, of neutrino and antineutrino,
are equal.

\subsection{Two-flavor oscillation with neutrino-antineutrino mixing}
In this case, the flavor and mass eigenstates are related via a unitary matrix $\textbf{T}$ as follows \cite{bm2}:
\begin{equation}
\begin{pmatrix}
\psi_e^c \\
\psi_\mu^c \\
\psi_e \\
\psi_\mu
\end{pmatrix}      =  \textbf{T}  \begin{pmatrix}
\chi_1 \\
\chi_2 \\
\chi_3 \\
\chi_4.
\end{pmatrix},
\end{equation}
where

\small
\begin{align}
\textbf{T} & \nonumber \\ =& \begin{pmatrix}
\cos\theta_e \cos\phi_1   &  - \cos\theta_e \sin\phi_1   &  -\sin\theta_e \cos\phi_2    & \sin\theta_e \sin\phi_2 \\
\cos\theta_\mu \sin\phi_1 &   \cos\theta_\mu \cos\phi_1  &  - \sin\theta_\mu \sin\phi_2 &  - \sin\theta_\mu \cos\phi_2 \\
\sin\theta_e \cos\phi_1   &   -\sin\theta_e \sin\phi_1   &  \cos\theta_e \cos\phi_2     & - \cos\theta_e \sin\phi_2 \\
\sin\theta_\mu \sin\phi_1 & \sin\theta_\mu \cos\phi_1    & \cos\theta_\mu \sin\phi_2    &  \cos\theta_\mu \cos\phi_2
\end{pmatrix}.
\end{align}
\normalsize

The mixing angles are related to the masses and the gravitational scalar potential as \cite{bm2}
\begin{align}
\tan \theta_{e, \mu} &= \frac{m_{e, \mu}}{B_0^g + \sqrt{(B_{0}^g)^2 + m_{e, \mu}^2}},\\
\tan \phi_{1,2}      &= \frac{\mp 2 m_{e\mu }}{ m_{e (1,2)} - m_{\mu (1,2)} + \sqrt{ (m_{e (1,2)} - m_{\mu (1,2)})^2 + 4 m_{e \mu}^2 }}.
\end{align}
The masses corresponding to the mass eigenstates are given as
\begin{align}
M_{1,2} &= \frac{1}{2} \big[ (m_{e1} + m_{\mu1}) \pm \sqrt{ (m_{e1} - m_{\mu 1})^2 + 4 m_{e \mu}^2 } \big], \nonumber\\
M_{3,4} &= \frac{1}{2} \big[ (m_{e2} + m_{\mu2}) \pm \sqrt{ (m_{e2} - m_{\mu2})^2 + 4 m_{e \mu}^2 } \big],
\end{align}
with 
\begin{equation}
m_{(e, \mu)1} = -\sqrt{(B_{0}^g)^2 + m_{e, \mu}^2} \qquad  m_{(e, \mu)2} = \sqrt{(B_{0}^g)^2 + m_{e, \mu}^2},
\end{equation}
and $m_{e\mu}$ being the mixing mass.

\begin{figure*}[ht] 
	\centering
		\includegraphics[width=150mm]{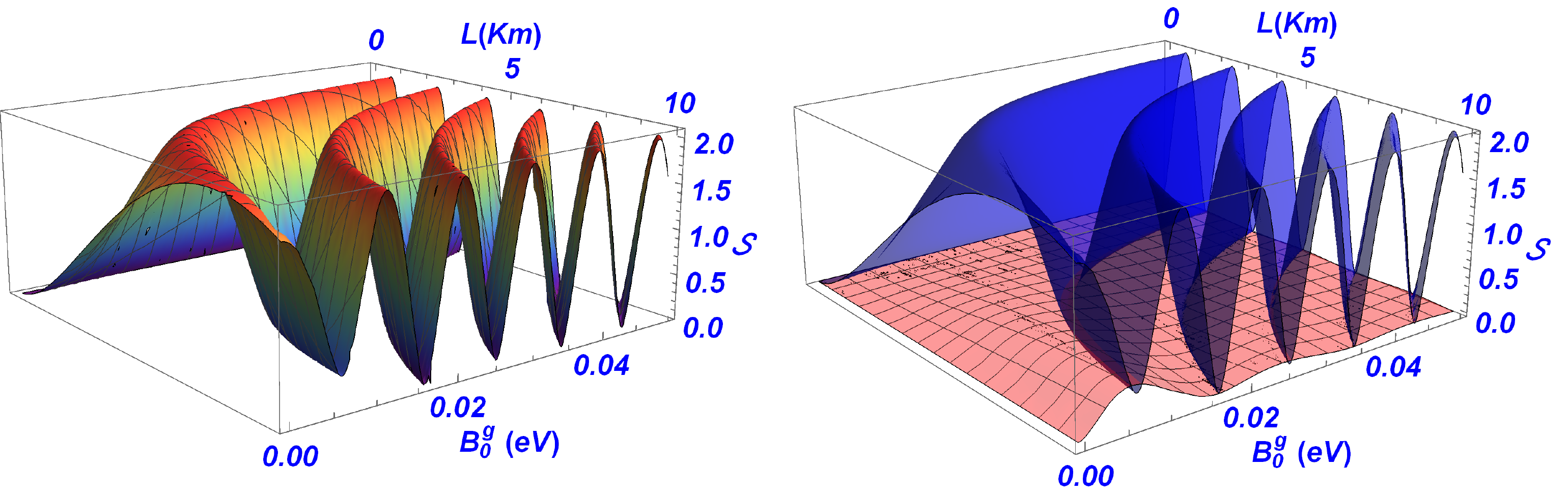}
	\caption{Left panel: The variation of von Neumann entropy for 2-flavor neutrino-antineutrino oscillation with respect to the distance ($L$) traveled by neutrinos and the gravitational potential ($B_0^g$). Right panel: The contributions to flavor entropy from neutrino-neutrino oscillations depicted by blue (plane) surface and neutrino-antineutrino oscillations  shown by pink (meshed) surface, with the magnitude of 
$\cal S$  enhanced  10 times in the later case.}
	\label{Von2}
\end{figure*}


\vskip0.5cm

Now the system of 2-flavor neutrino oscillations under the influence of neutrino-antineutrino mixing, due to gravitational field, can be treated as a 4-qubit system. The occupation number representation can be given as
\begin{eqnarray}\label{occupation2}
\ket{\psi_e^c} \equiv \ket{1}_{\bar{e}}\otimes\ket{0}_{\bar{\mu}}\otimes\ket{0}_{e}\otimes\ket{0}_{\mu},\\  \nonumber
\ket{\psi_{\mu}^c} \equiv \ket{0}_{\bar{e}}\otimes\ket{1}_{\bar{\mu}}\otimes\ket{0}_{e}\otimes\ket{0}_{\mu},\\ \nonumber
\ket{\psi_e} \equiv \ket{0}_{\bar{e}}\otimes\ket{0}_{\bar{\mu}}\otimes\ket{1}_{e}\otimes\ket{0}_{\mu},\\ \nonumber
\ket{\psi_{\mu}} \equiv \ket{0}_{\bar{e}}\otimes\ket{0}_{\bar{\mu}}\otimes\ket{0}_{e}\otimes\ket{1}_{\mu}. 
\end{eqnarray}

Further, Mermin inequality is a generalized form of Bell inequality and its violation indicates the standard nonlocal correlations existing among different parties in a multipartite system \cite{Mermin}. This means that the probability distribution $P$ (say for a tripartite system) cannot be written in the local form
\begin{equation}
P(a_1 a_2 a_3) = \int d\lambda \rho(\lambda) P_1(a_1|\lambda) P_2(a_2|\lambda) P_3(a_3|\lambda),\label{probability}
\end{equation}
where $\lambda$ is the shared local variable and $a_1$, $a_2$, $a_3$ are the outcomes of the measurements. However, this does not ensure the genuine multipartite nonlocality, i.e., if any two subsystems are nonlocally correlated, but uncorrelated from the third one, Mermin inequality can  still be violated \cite{Collins,Gisin}. To probe genuine nonlocal correlations, we make use of the Svetlichny inequality which is based on hybrid nonlocal-local realism \cite{Svetlichny} as follows
\begin{equation}
P_B(a_1 a_2 a_3) = \sum_{k=1}^{3} P_k \int d\lambda \rho_{ij}(\lambda) P_{ij}(a_i a_j|\lambda) P_k(a_k|\lambda).\label{probability2}
\end{equation}
Here the subscript $B$ stands for bipartition sections.  
For a 4-qubit-system the Mermin ($M_4$) \cite{Alsina} and Svetlichny ($S_4$) \cite{Bancal} parameters are defined as
\begin{align}
M_4 & = -ABCD + (ABCD' + ABC'D + AB'CD + A'BCD) \nonumber\\
& + (ABC'D' + AB'CD' + AB'C'D + A'BCD' + A'BC'D \nonumber\\ 
& + A'B'CD) - (AB'C'D' + A'BC'D' + A'B'CD' \nonumber\\
& + A'B'C'D) - A'B'C'D',
\end{align}
\begin{align}
S_4 & = ABC'D' + AB'CD' + A'BCD' - A'B'C'D' + \nonumber\\
& A'B'CD' + A'BC'D' + AB'C'D' - AB'C'D' + A'B'CD \nonumber\\
& + A'BC'D + AB'C'D - ABCD + ABC'D + AB'CD + \nonumber\\ 
& A'BCD - A'B'C'D.
\end{align}
Here,  $X$ and $X^\prime$ ($X=A, B, C, D$), are two different measurement settings pertaining to each qubit.
The classical bounds on these parameters are $M_4 \leq 4$ and $S_4 \leq 8$.  It is important to note that for the violation of Mermin inequality, at least one bipartite section must have the nonlocal correlations, while the Svetlichny inequality will be violated only when all the parties are nonlocally correlated.
\begin{figure}
	\centering
	(a)\includegraphics[scale=0.46]{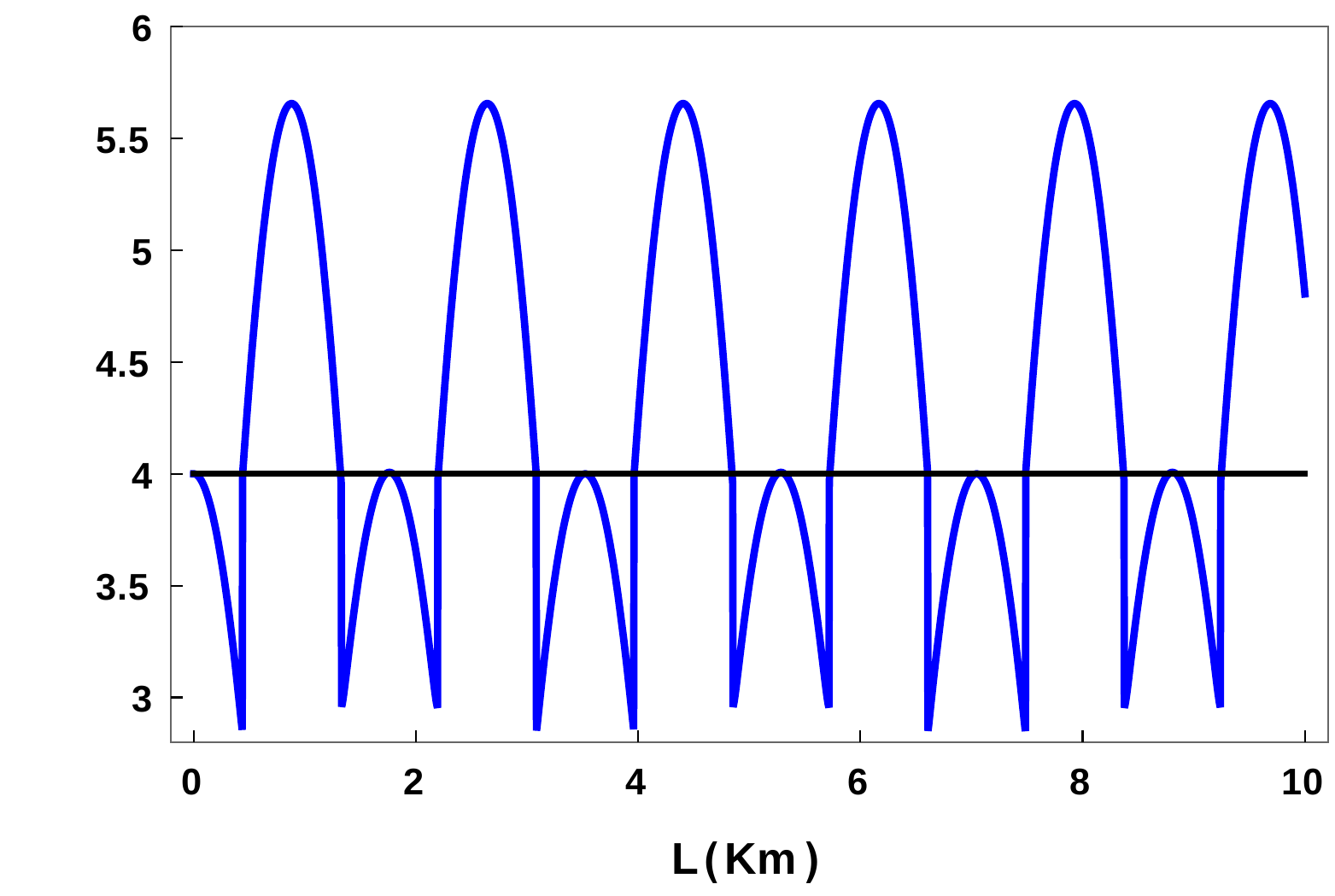} 
	(b)\includegraphics[scale=0.46]{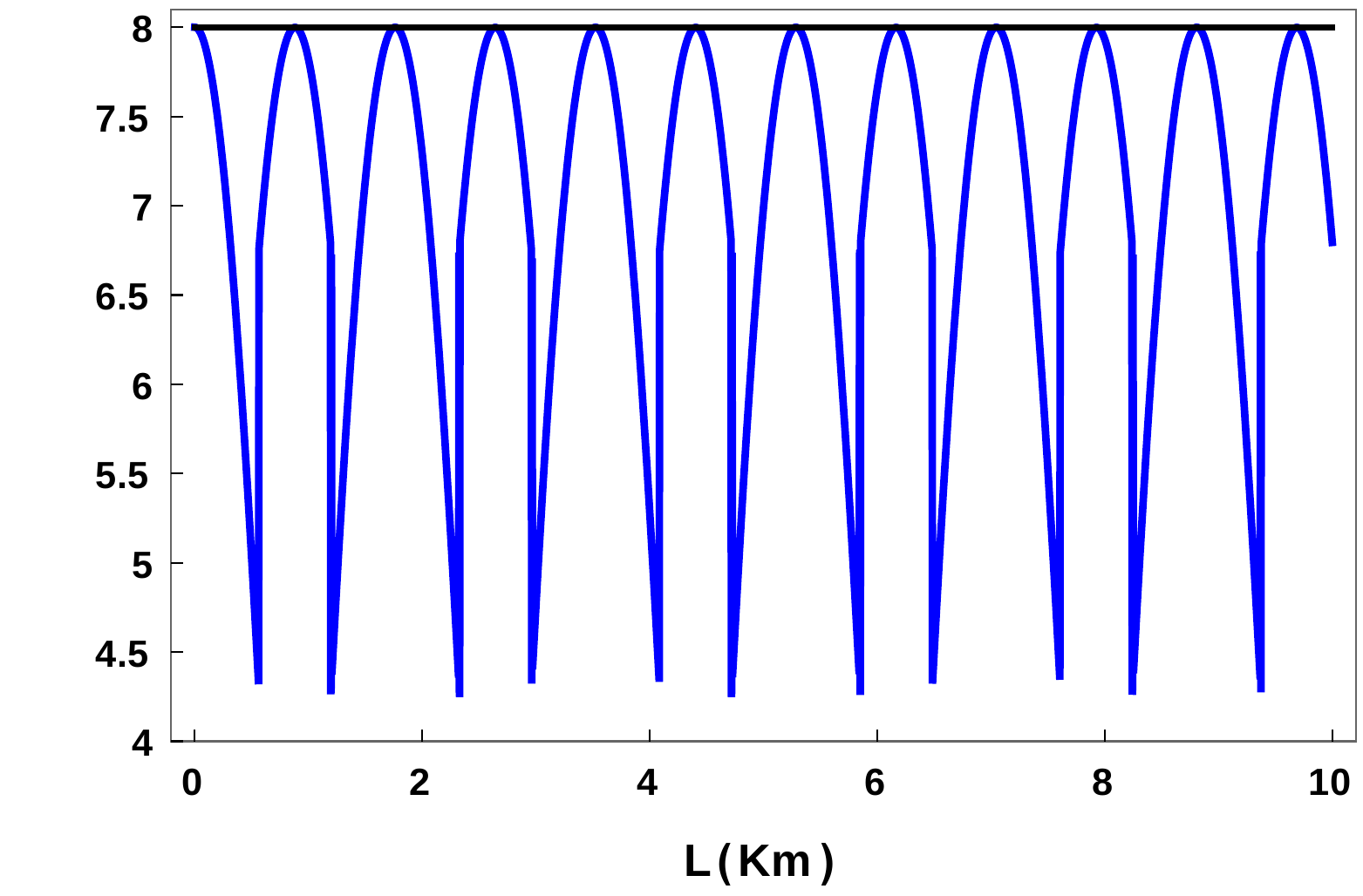}
	\caption{(a) Mermin $M_4$ and (b) Svetlichny $S_4$ parameters with respect to the distance ($L$) traveled by neutrinos with $B_0^g = 5\times10^{-2}$. Black lines correspond to the classical bounds of these parameters.}
	\label{nonlocality}
\end{figure}
\FloatBarrier

In Fig. \ref{Von2}, the variation of von-Neumann entropy ($\mathcal{S}$) is depicted with respect to the distance $L$ traveled by neutrinos and the gravitational potential $B_0^g$. In two flavor case the state of the system has four degrees of freedom to oscillate between. With initial state as $\nu_e$, oscillations occur between $\nu_e$, $\nu_{\mu}$, $\bar{\nu}_e$ and $\bar{\nu}_{\mu}$ flavor modes of the system. Figure \ref{Von2}a depicts the total flavor entropy with contribution from all the available modes of oscillation, while in Fig. \ref{Von2}b the contribution from particle and antiparticle modes separately is shown. For the sake of clarity, we have enhanced the magnitude of $\mathcal{S}$ by 10 times for antineutrino case.  
	The particle degrees of freedom contribute more to $\mathcal{S}$ in comparison to the antiparticle degrees of freedom. That is, the neutrino-neutrino flavor mixing is dominating over the neutrino-antineutrino mixing. A common feature depicted in Figs. \ref{Von} and \ref{Von2} is that for neutrino-antineutrino oscillations,  $\mathcal{S}$ decreases with the increase in $B_0^g$. However, for neutrino-neutrino mixing, the increase in $B_0^g$ does not reduce the magnitude of $\mathcal{S}$ but increases the frequency of its oscillation. Figure \ref{nonlocality} depicts Mermin and Svetlichny parameters with respect to $L$, with $B_0^g = 5 \times 10^{-2}$ eV. The violation of the classical bound of $M_4$ indicates the existence of residual nonlocality in the system. Further, $S_4$ does not cross the classical bound in our system, thereby showing the absence of genuine nonlocality. This can be attributed to equality of the two mass squared differences, i.e., $\Delta_{41} = \Delta_{32} = 0$, suppressing the nonlocal correlations between the degenerate levels $\nu_1-\nu_4$ and $\nu_2-\nu_3$. 

\section{Summary and Conclusion}
Spinors interacting with background gravitational field in an arbitrary spacetime are shown to acquire modified dispersion energy with energies for up and down spinors split: gravitational Zeeman splitting. It has important consequences for neutrino mixing and oscillation and in general various quantum correlations. However, for this to have occurred, at least in local coordinates, the spacetime should not be spherical symmetric. To have a nontrivial oscillation phase induced by the gravitational field of, e.g., black holes, such that effects due to neutrino mass are not dominant, the mass of the black hole producing gravitational fields must not be more than a millionth of a solar mass, hence be primordial in nature. In other words, only primordial black holes give rise to any practical gravitational Zeeman splitting. However, without gravitational Zeeman splitting there is no neutrino-antineutrino oscillation even if mixing is nonzero. Moreover, apart from black hole, there are many other scenarios, for e.g., early universe particularly in presence of tensor/gravitational perturbation, where strong gravitational effect leads to neutrino-antineutrino mixing and oscillation. 

There are interesting consequences of the interplay between the influence of gravity on neutrino-antineutrino oscillations. Thus, for example, for a single flavor neutrino-antineutrino oscillations, entanglement is maximum for the case when the neutrino and antineutrino states are equally probable. Gravity however suppresses the entanglement between neutrino and antineutrino states, which is implied by a decrease in the von-Neumann entropy $\mathcal{S}$ with the increase in the gravitation potential $B_0^g$.
	
Further, in case of 2-flavor neutrino-antineutrino oscillations, $\mathcal{S}$ is non-zero which indicates absolute entanglement in the system. Mermin inequality is violated while Svetlichny is not, implying that the system is having absolute nonlocal correlations (nonlocality shared by at least two parties) but  genuine nonlocal correlations (nonlocality shared among all parties) are absent, a consequence of the degeneracy in the levels $\nu_1-\nu_4$ and $\nu_2-\nu_3$.


\begin{thebibliography}{52}%
	\makeatletter
	\providecommand \@ifxundefined [1]{%
		\@ifx{#1\undefined}
	}%
	\providecommand \@ifnum [1]{%
		\ifnum #1\expandafter \@firstoftwo
		\else \expandafter \@secondoftwo
		\fi
	}%
	\providecommand \@ifx [1]{%
		\ifx #1\expandafter \@firstoftwo
		\else \expandafter \@secondoftwo
		\fi
	}%
	\providecommand \natexlab [1]{#1}%
	\providecommand \enquote  [1]{``#1''}%
	\providecommand \bibnamefont  [1]{#1}%
	\providecommand \bibfnamefont [1]{#1}%
	\providecommand \citenamefont [1]{#1}%
	\providecommand \href@noop [0]{\@secondoftwo}%
	\providecommand \href [0]{\begingroup \@sanitize@url \@href}%
	\providecommand \@href[1]{\@@startlink{#1}\@@href}%
	\providecommand \@@href[1]{\endgroup#1\@@endlink}%
	\providecommand \@sanitize@url [0]{\catcode `\\12\catcode `\$12\catcode
		`\&12\catcode `\#12\catcode `\^12\catcode `\_12\catcode `\%12\relax}%
	\providecommand \@@startlink[1]{}%
	\providecommand \@@endlink[0]{}%
	\providecommand \url  [0]{\begingroup\@sanitize@url \@url }%
	\providecommand \@url [1]{\endgroup\@href {#1}{\urlprefix }}%
	\providecommand \urlprefix  [0]{URL }%
	\providecommand \Eprint [0]{\href }%
	\providecommand \doibase [0]{http://dx.doi.org/}%
	\providecommand \selectlanguage [0]{\@gobble}%
	\providecommand \bibinfo  [0]{\@secondoftwo}%
	\providecommand \bibfield  [0]{\@secondoftwo}%
	\providecommand \translation [1]{[#1]}%
	\providecommand \BibitemOpen [0]{}%
	\providecommand \bibitemStop [0]{}%
	\providecommand \bibitemNoStop [0]{.\EOS\space}%
	\providecommand \EOS [0]{\spacefactor3000\relax}%
	\providecommand \BibitemShut  [1]{\csname bibitem#1\endcsname}%
	\let\auto@bib@innerbib\@empty
	\bibitem [{Sup(2003)}]{SuperK}%
	\BibitemOpen
	\bibfield  {author} {\bibinfo {author} {\bibfnamefont {S.}\
			\bibnamefont {Fukuda {\it et al.}}},\ }
	\href {\doibase https://doi.org/10.1016/S0168-9002(03)00425-X} {\bibfield
		{journal} {\bibinfo  {journal} {Nucl. Instrum. Methods Phys.
				Res. Sect. A}\ }\textbf {\bibinfo {volume} {501}},\ \bibinfo {pages} {418 }
		(\bibinfo {year} {2003})}\BibitemShut {NoStop}%
	\bibitem [{\citenamefont {Ahmad}(2002)}]{SNO}%
	\BibitemOpen
	\bibfield  {author} {\bibinfo {author} {\bibfnamefont {Q.~R. e.~a.}\
			\bibnamefont {Ahmad}} (\bibinfo {collaboration} {SNO Collaboration}),\ }\href
	{\doibase 10.1103/PhysRevLett.89.011301} {\bibfield  {journal} {\bibinfo
			{journal} {Physical Review Letters}\ }\textbf {\bibinfo {volume} {89}},\
		\bibinfo {pages} {011301} (\bibinfo {year} {2002})}\BibitemShut {NoStop}%
	\bibitem [{\citenamefont {Schechter}\ and\ \citenamefont
		{Valle}(1981)}]{Schechter1981}%
	\BibitemOpen
	\bibfield  {author} {\bibinfo {author} {\bibfnamefont {J.}~\bibnamefont
			{Schechter}}\ and\ \bibinfo {author} {\bibfnamefont {J.~W.~F.}\ \bibnamefont
			{Valle}},\ }\href {\doibase 10.1103/PhysRevD.23.1666} {\bibfield  {journal}
		{\bibinfo  {journal} {Phys. Rev. D}\ }\textbf {\bibinfo {volume} {23}},\
		\bibinfo {pages} {1666} (\bibinfo {year} {1981})}\BibitemShut {NoStop}%
	\bibitem [{\citenamefont {Kobzarev}\ \emph {et~al.}(1981)\citenamefont
		{Kobzarev}, \citenamefont {Martemyanov}, \citenamefont {Okun},\ and\
		\citenamefont {Schepkin}}]{kobzarev1981sum}%
	\BibitemOpen
	\bibfield  {author} {\bibinfo {author} {\bibfnamefont {I.~Y.}\ \bibnamefont
			{Kobzarev}}, \bibinfo {author} {\bibfnamefont {B.}~\bibnamefont
			{Martemyanov}}, \bibinfo {author} {\bibfnamefont {L.}~\bibnamefont {Okun}}, \
		and\ \bibinfo {author} {\bibfnamefont {M.}~\bibnamefont {Schepkin}},\
	}\href@noop {} {\emph {\bibinfo {title} {Sum rules for neutrino
				oscillations}}},\ \bibinfo {type} {Tech. Rep.}\ (\bibinfo  {institution}
	{Gosudarstvennyj Komitet po Ispol'zovaniyu Atomnoj Ehnergii SSSR},\ \bibinfo
	{year} {1981})\BibitemShut {NoStop}%
	\bibitem [{\citenamefont {Sinha}\ and\ \citenamefont
		{Mukhopadhyay}(2008)}]{bm2}%
	\BibitemOpen
	\bibfield  {author} {\bibinfo {author} {\bibfnamefont {M.}~\bibnamefont
			{Sinha}}\ and\ \bibinfo {author} {\bibfnamefont {B.}~\bibnamefont
			{Mukhopadhyay}},\ }\href {\doibase 10.1103/PhysRevD.77.025003} {\bibfield
		{journal} {\bibinfo  {journal} {Physical Review D}\ }\textbf {\bibinfo
			{volume} {77}},\ \bibinfo {pages} {025003} (\bibinfo {year}
		{2008})}\BibitemShut {NoStop}%
	\bibitem [{\citenamefont {Mukhopadhyay}(2007)}]{bm1}%
	\BibitemOpen
	\bibfield  {author} {\bibinfo {author} {\bibfnamefont {B.}~\bibnamefont
			{Mukhopadhyay}},\ }\href@noop {} {\bibfield  {journal} {\bibinfo  {journal}
			{Classical and Quantum Gravity}\ }\textbf {\bibinfo {volume} {24}},\ \bibinfo
		{pages} {1433} (\bibinfo {year} {2007})}\BibitemShut {NoStop}%
	\bibitem [{\citenamefont {Mohanty}\ \emph {et~al.}(2002)\citenamefont
		{Mohanty}, \citenamefont {Mukhopadhyay},\ and\ \citenamefont
		{Prasanna}}]{bmprd02}%
	\BibitemOpen
	\bibfield  {author} {\bibinfo {author} {\bibfnamefont {S.}~\bibnamefont
			{Mohanty}}, \bibinfo {author} {\bibfnamefont {B.}~\bibnamefont
			{Mukhopadhyay}}, \ and\ \bibinfo {author} {\bibfnamefont {A.~R.}\
			\bibnamefont {Prasanna}},\ }\href {\doibase 10.1103/PhysRevD.65.122001}
	{\bibfield  {journal} {\bibinfo  {journal} {Physical Review D}\ }\textbf
		{\bibinfo {volume} {65}},\ \bibinfo {pages} {122001} (\bibinfo {year}
		{2002})}\BibitemShut {NoStop}%
	\bibitem [{\citenamefont {Mukhopadhyay}(2005)}]{bmmpla05}%
	\BibitemOpen
	\bibfield  {author} {\bibinfo {author} {\bibfnamefont {B.}~\bibnamefont
			{Mukhopadhyay}},\ }\href {\doibase 10.1142/S0217732305017640} {\bibfield
		{journal} {\bibinfo  {journal} {Modern Physics Letters A}\ }\textbf {\bibinfo
			{volume} {20}},\ \bibinfo {pages} {2145} (\bibinfo {year}
		{2005})}\BibitemShut {NoStop}%
	\bibitem [{\citenamefont {Debnath}\ \emph {et~al.}(2006)\citenamefont
		{Debnath}, \citenamefont {Mukhopadhyay},\ and\ \citenamefont
		{Dadhich}}]{bmujjal06}%
	\BibitemOpen
	\bibfield  {author} {\bibinfo {author} {\bibfnamefont {U.}~\bibnamefont
			{Debnath}}, \bibinfo {author} {\bibfnamefont {B.}~\bibnamefont
			{Mukhopadhyay}}, \ and\ \bibinfo {author} {\bibfnamefont {N.}~\bibnamefont
			{Dadhich}},\ }\href {\doibase 10.1142/S0217732306019542} {\bibfield
		{journal} {\bibinfo  {journal} {Modern Physics Letters A}\ }\textbf {\bibinfo
			{volume} {21}},\ \bibinfo {pages} {399} (\bibinfo {year} {2006})}\BibitemShut
	{NoStop}%
	\bibitem [{\citenamefont {Mavromatos}\ and\ \citenamefont
		{Sarkar}(2013)}]{mavromatosepjc13}%
	\BibitemOpen
	\bibfield  {author} {\bibinfo {author} {\bibfnamefont {N.~E.}\ \bibnamefont
			{Mavromatos}}\ and\ \bibinfo {author} {\bibfnamefont {S.}~\bibnamefont
			{Sarkar}},\ }\href@noop {} {\bibfield  {journal} {\bibinfo  {journal} {The
				European Physical Journal C}\ }\textbf {\bibinfo {volume} {73}},\ \bibinfo
		{pages} {2359} (\bibinfo {year} {2013})}\BibitemShut {NoStop}%
	\bibitem [{\citenamefont {Cuesta}(2017)}]{mosqueraapj17}%
	\BibitemOpen
	\bibfield  {author} {\bibinfo {author} {\bibfnamefont {H.~J.~M.}\
			\bibnamefont {Cuesta}},\ }\href@noop {} {\bibfield  {journal} {\bibinfo
			{journal} {The Astrophysical Journal}\ }\textbf {\bibinfo {volume} {835}},\
		\bibinfo {pages} {215} (\bibinfo {year} {2017})}\BibitemShut {NoStop}%
	\bibitem [{\citenamefont {Blasone}\ \emph {et~al.}(1999)\citenamefont
		{Blasone}, \citenamefont {Henning},\ and\ \citenamefont
		{Vitiello}}]{BLASONE1999}%
	\BibitemOpen
	\bibfield  {author} {\bibinfo {author} {\bibfnamefont {M.}~\bibnamefont
			{Blasone}}, \bibinfo {author} {\bibfnamefont {P.~A.}\ \bibnamefont
			{Henning}}, \ and\ \bibinfo {author} {\bibfnamefont {G.}~\bibnamefont
			{Vitiello}},\ }\href {\doibase https://doi.org/10.1016/S0370-2693(99)01137-5}
	{\bibfield  {journal} {\bibinfo  {journal} {Physics Letters B}\ }\textbf
		{\bibinfo {volume} {466}},\ \bibinfo {pages} {262 } (\bibinfo {year}
		{1999})}\BibitemShut {NoStop}%
	\bibitem [{\citenamefont {He}\ \emph {et~al.}(2005)\citenamefont {He},
		\citenamefont {Li}, \citenamefont {McKellar},\ and\ \citenamefont
		{Zhang}}]{He2005}%
	\BibitemOpen
	\bibfield  {author} {\bibinfo {author} {\bibfnamefont {X.-G.}\ \bibnamefont
			{He}}, \bibinfo {author} {\bibfnamefont {X.-Q.}\ \bibnamefont {Li}}, \bibinfo
		{author} {\bibfnamefont {B.~H.~J.}\ \bibnamefont {McKellar}}, \ and\ \bibinfo
		{author} {\bibfnamefont {Y.}~\bibnamefont {Zhang}},\ }\href {\doibase
		10.1103/PhysRevD.72.053012} {\bibfield  {journal} {\bibinfo  {journal}
			{Physical Review D}\ }\textbf {\bibinfo {volume} {72}},\ \bibinfo {pages}
		{053012} (\bibinfo {year} {2005})}\BibitemShut {NoStop}%
	\bibitem [{\citenamefont {Wang}\ \emph {et~al.}(2001)\citenamefont {Wang},
		\citenamefont {Kwek}, \citenamefont {Liu},\ and\ \citenamefont
		{Oh}}]{Wang2001}%
	\BibitemOpen
	\bibfield  {author} {\bibinfo {author} {\bibfnamefont {X.-B.}\ \bibnamefont
			{Wang}}, \bibinfo {author} {\bibfnamefont {L.~C.}\ \bibnamefont {Kwek}},
		\bibinfo {author} {\bibfnamefont {Y.}~\bibnamefont {Liu}}, \ and\ \bibinfo
		{author} {\bibfnamefont {C.~H.}\ \bibnamefont {Oh}},\ }\href {\doibase
		10.1103/PhysRevD.63.053003} {\bibfield  {journal} {\bibinfo  {journal}
			{Physical Review D}\ }\textbf {\bibinfo {volume} {63}},\ \bibinfo {pages}
		{053003} (\bibinfo {year} {2001})}\BibitemShut {NoStop}%
	\bibitem [{\citenamefont {Dajka}\ \emph {et~al.}(2011)\citenamefont {Dajka},
		\citenamefont {Syska},\ and\ \citenamefont {\L{}uczka}}]{Dajka2011}%
	\BibitemOpen
	\bibfield  {author} {\bibinfo {author} {\bibfnamefont {J.}~\bibnamefont
			{Dajka}}, \bibinfo {author} {\bibfnamefont {J.}~\bibnamefont {Syska}}, \ and\
		\bibinfo {author} {\bibfnamefont {J.}~\bibnamefont {\L{}uczka}},\ }\href
	{\doibase 10.1103/PhysRevD.83.097302} {\bibfield  {journal} {\bibinfo
			{journal} {Physical Review D}\ }\textbf {\bibinfo {volume} {83}},\ \bibinfo
		{pages} {097302} (\bibinfo {year} {2011})}\BibitemShut {NoStop}%
	\bibitem [{\citenamefont {Mehta}(2009)}]{Mehta2009}%
	\BibitemOpen
	\bibfield  {author} {\bibinfo {author} {\bibfnamefont {P.}~\bibnamefont
			{Mehta}},\ }\href {\doibase 10.1103/PhysRevD.79.096013} {\bibfield  {journal}
		{\bibinfo  {journal} {Physical Review D}\ }\textbf {\bibinfo {volume} {79}},\
		\bibinfo {pages} {096013} (\bibinfo {year} {2009})}\BibitemShut {NoStop}%
	\bibitem [{\citenamefont {Dixit}\ \emph
		{et~al.}(2018{\natexlab{a}})\citenamefont {Dixit}, \citenamefont {Alok},
		\citenamefont {Banerjee},\ and\ \citenamefont {Kumar}}]{Dixit_2018}%
	\BibitemOpen
	\bibfield  {author} {\bibinfo {author} {\bibfnamefont {K.}~\bibnamefont
			{Dixit}}, \bibinfo {author} {\bibfnamefont {A.~K.}\ \bibnamefont {Alok}},
		\bibinfo {author} {\bibfnamefont {S.}~\bibnamefont {Banerjee}}, \ and\
		\bibinfo {author} {\bibfnamefont {D.}~\bibnamefont {Kumar}},\ }\href
	{\doibase 10.1088/1361-6471/aac454} {\bibfield  {journal} {\bibinfo
			{journal} {Journal of Physics G: Nuclear and Particle Physics}\ }\textbf
		{\bibinfo {volume} {45}},\ \bibinfo {pages} {085002} (\bibinfo {year}
		{2018}{\natexlab{a}})}\BibitemShut {NoStop}%
	\bibitem [{\citenamefont {Joshi}\ and\ \citenamefont {Jain}(2016)}]{joshi2016}%
	\BibitemOpen
	\bibfield  {author} {\bibinfo {author} {\bibfnamefont {S.}~\bibnamefont
			{Joshi}}\ and\ \bibinfo {author} {\bibfnamefont {S.~R.}\ \bibnamefont
			{Jain}},\ }\href@noop {} {\bibfield  {journal} {\bibinfo  {journal} {Physics
				Letters B}\ }\textbf {\bibinfo {volume} {754}},\ \bibinfo {pages} {135}
		(\bibinfo {year} {2016})}\BibitemShut {NoStop}%
	\bibitem [{\citenamefont {Mukhopadhyay}\ and\ \citenamefont
		{Ganguly}()}]{bm2018}%
	\BibitemOpen
	\bibfield  {author} {\bibinfo {author} {\bibfnamefont {B.}~\bibnamefont
			{Mukhopadhyay}}\ and\ \bibinfo {author} {\bibfnamefont {S.~K.}\ \bibnamefont
			{Ganguly}},\ }\href@noop {} {\bibinfo  {journal} {arXiv:1802.10377 [gr-qc]}\
	}\BibitemShut {NoStop}%
	\bibitem [{\citenamefont {Bell}(1964)}]{Bell}%
	\BibitemOpen
	\bibfield  {journal} {  }\bibfield  {author} {\bibinfo {author} {\bibfnamefont
			{J.~S.}\ \bibnamefont {Bell}},\ }\href@noop {} {\bibfield  {journal}
		{\bibinfo  {journal} {Physics Physique Fizika}\ }\textbf {\bibinfo {volume}
			{1}},\ \bibinfo {pages} {195} (\bibinfo {year} {1964})}\BibitemShut {NoStop}%
	\bibitem [{\citenamefont {Aspect}\ \emph {et~al.}(1981)\citenamefont {Aspect},
		\citenamefont {Grangier},\ and\ \citenamefont {Roger}}]{Aspect}%
	\BibitemOpen
	\bibfield  {author} {\bibinfo {author} {\bibfnamefont {A.}~\bibnamefont
			{Aspect}}, \bibinfo {author} {\bibfnamefont {P.}~\bibnamefont {Grangier}}, \
		and\ \bibinfo {author} {\bibfnamefont {G.}~\bibnamefont {Roger}},\ }\href
	{\doibase 10.1103/PhysRevLett.47.460} {\bibfield  {journal} {\bibinfo
			{journal} {Physical Review Letters}\ }\textbf {\bibinfo {volume} {47}},\
		\bibinfo {pages} {460} (\bibinfo {year} {1981})}\BibitemShut {NoStop}%
	\bibitem [{\citenamefont {Tittel}\ \emph {et~al.}(1998)\citenamefont {Tittel},
		\citenamefont {Brendel}, \citenamefont {Gisin}, \citenamefont {Herzog},
		\citenamefont {Zbinden},\ and\ \citenamefont {Gisin}}]{Tittel}%
	\BibitemOpen
	\bibfield  {author} {\bibinfo {author} {\bibfnamefont {W.}~\bibnamefont
			{Tittel}}, \bibinfo {author} {\bibfnamefont {J.}~\bibnamefont {Brendel}},
		\bibinfo {author} {\bibfnamefont {B.}~\bibnamefont {Gisin}}, \bibinfo
		{author} {\bibfnamefont {T.}~\bibnamefont {Herzog}}, \bibinfo {author}
		{\bibfnamefont {H.}~\bibnamefont {Zbinden}}, \ and\ \bibinfo {author}
		{\bibfnamefont {N.}~\bibnamefont {Gisin}},\ }\href {\doibase
		10.1103/PhysRevA.57.3229} {\bibfield  {journal} {\bibinfo  {journal}
			{Physical Review A}\ }\textbf {\bibinfo {volume} {57}},\ \bibinfo {pages}
		{3229} (\bibinfo {year} {1998})}\BibitemShut {NoStop}%
	\bibitem [{\citenamefont {Lanyon}\ \emph {et~al.}(2013)\citenamefont {Lanyon},
		\citenamefont {Jurcevic}, \citenamefont {Hempel}, \citenamefont {Gessner},
		\citenamefont {Vedral}, \citenamefont {Blatt},\ and\ \citenamefont
		{Roos}}]{Lanyon}%
	\BibitemOpen
	\bibfield  {author} {\bibinfo {author} {\bibfnamefont {B.~P.}\ \bibnamefont
			{Lanyon}}, \bibinfo {author} {\bibfnamefont {P.}~\bibnamefont {Jurcevic}},
		\bibinfo {author} {\bibfnamefont {C.}~\bibnamefont {Hempel}}, \bibinfo
		{author} {\bibfnamefont {M.}~\bibnamefont {Gessner}}, \bibinfo {author}
		{\bibfnamefont {V.}~\bibnamefont {Vedral}}, \bibinfo {author} {\bibfnamefont
			{R.}~\bibnamefont {Blatt}}, \ and\ \bibinfo {author} {\bibfnamefont {C.~F.}\
			\bibnamefont {Roos}},\ }\href {\doibase 10.1103/PhysRevLett.111.100504}
	{\bibfield  {journal} {\bibinfo  {journal} {Physical Review Letters}\
		}\textbf {\bibinfo {volume} {111}},\ \bibinfo {pages} {100504} (\bibinfo
		{year} {2013})}\BibitemShut {NoStop}%
	\bibitem [{\citenamefont {Chakrabarty}\ \emph {et~al.}(2011)\citenamefont
		{Chakrabarty}, \citenamefont {Banerjee},\ and\ \citenamefont
		{Siddharth}}]{Chakrabarty}%
	\BibitemOpen
	\bibfield  {author} {\bibinfo {author} {\bibfnamefont {I.}~\bibnamefont
			{Chakrabarty}}, \bibinfo {author} {\bibfnamefont {S.}~\bibnamefont
			{Banerjee}}, \ and\ \bibinfo {author} {\bibfnamefont {N.}~\bibnamefont
			{Siddharth}},\ }\href@noop {} {\bibfield  {journal} {\bibinfo  {journal}
			{Quantum Inf. Comput.}\ }\textbf {\bibinfo {volume} {11}},\ \bibinfo {pages}
		{0541} (\bibinfo {year} {2011})}\BibitemShut {NoStop}%
	\bibitem [{\citenamefont {Banerjee}\ \emph {et~al.}(2016)\citenamefont
		{Banerjee}, \citenamefont {Alok},\ and\ \citenamefont
		{MacKenzie}}]{MacKenzie}%
	\BibitemOpen
	\bibfield  {author} {\bibinfo {author} {\bibfnamefont {S.}~\bibnamefont
			{Banerjee}}, \bibinfo {author} {\bibfnamefont {A.~K.}\ \bibnamefont {Alok}},
		\ and\ \bibinfo {author} {\bibfnamefont {R.}~\bibnamefont {MacKenzie}},\
	}\href@noop {} {\bibfield  {journal} {\bibinfo  {journal} {The European
				Physical Journal Plus}\ }\textbf {\bibinfo {volume} {131}},\ \bibinfo {pages}
		{129} (\bibinfo {year} {2016})}\BibitemShut {NoStop}%
	\bibitem [{\citenamefont {Naikoo}\ \emph {et~al.}(2018)\citenamefont {Naikoo},
		\citenamefont {Alok},\ and\ \citenamefont {Banerjee}}]{Naikoo1}%
	\BibitemOpen
	\bibfield  {author} {\bibinfo {author} {\bibfnamefont {J.}~\bibnamefont
			{Naikoo}}, \bibinfo {author} {\bibfnamefont {A.~K.}\ \bibnamefont {Alok}}, \
		and\ \bibinfo {author} {\bibfnamefont {S.}~\bibnamefont {Banerjee}},\ }\href
	{\doibase 10.1103/PhysRevD.97.053008} {\bibfield  {journal} {\bibinfo
			{journal} {Physical Review D}\ }\textbf {\bibinfo {volume} {97}},\ \bibinfo
		{pages} {053008} (\bibinfo {year} {2018})}\BibitemShut {NoStop}%
	\bibitem [{\citenamefont {Caban}\ \emph {et~al.}(2006)\citenamefont {Caban},
		\citenamefont {Rembieli{\'n}ski}, \citenamefont {Smoli{\'n}ski},
		\citenamefont {Walczak},\ and\ \citenamefont {W{\l}odarczyk}}]{Caban}%
	\BibitemOpen
	\bibfield  {author} {\bibinfo {author} {\bibfnamefont {P.}~\bibnamefont
			{Caban}}, \bibinfo {author} {\bibfnamefont {J.}~\bibnamefont
			{Rembieli{\'n}ski}}, \bibinfo {author} {\bibfnamefont {K.~A.}\ \bibnamefont
			{Smoli{\'n}ski}}, \bibinfo {author} {\bibfnamefont {Z.}~\bibnamefont
			{Walczak}}, \ and\ \bibinfo {author} {\bibfnamefont {M.}~\bibnamefont
			{W{\l}odarczyk}},\ }\href@noop {} {\bibfield  {journal} {\bibinfo  {journal}
			{Physics Letters A}\ }\textbf {\bibinfo {volume} {357}},\ \bibinfo {pages}
		{6} (\bibinfo {year} {2006})}\BibitemShut {NoStop}%
	\bibitem [{\citenamefont {Bramon}\ \emph {et~al.}(2007)\citenamefont {Bramon},
		\citenamefont {Garbarino},\ and\ \citenamefont {Escribano}}]{Bramon}%
	\BibitemOpen
	\bibfield  {author} {\bibinfo {author} {\bibfnamefont {A.}~\bibnamefont
			{Bramon}}, \bibinfo {author} {\bibfnamefont {R.}~\bibnamefont {Escribano}}, \
		and\ \bibinfo {author} {\bibfnamefont {G.}~\bibnamefont {Garbarino}},\
	}\href@noop {}  {\bibinfo {title} {A review of
			Bell inequality tests with neutral kaons, in \textit{Handbook
			on Neutral Kaon Interferometry at a $\Phi$-factory}}} {\bibfield  {journal} {\bibinfo  {editor} {edited by A. Di Domenico},\ }\textbf {\bibinfo {volume} {XLIII}},\ \bibinfo {pages} {217-254} (\bibinfo
		{year} {2007})}\BibitemShut {NoStop}%
	\bibitem [{\citenamefont {Alok}\ \emph {et~al.}(2016)\citenamefont {Alok},
		\citenamefont {Banerjee},\ and\ \citenamefont {Sankar}}]{Alok}%
	\BibitemOpen
	\bibfield  {author} {\bibinfo {author} {\bibfnamefont {A.~K.}\ \bibnamefont
			{Alok}}, \bibinfo {author} {\bibfnamefont {S.}~\bibnamefont {Banerjee}}, \
		and\ \bibinfo {author} {\bibfnamefont {S.~U.}\ \bibnamefont {Sankar}},\
	}\href@noop {} {\bibfield  {journal} {\bibinfo  {journal} {Nuclear Physics
				B}\ }\textbf {\bibinfo {volume} {909}},\ \bibinfo {pages} {65} (\bibinfo
		{year} {2016})}\BibitemShut {NoStop}%
	\bibitem [{\citenamefont {Banerjee}\ \emph {et~al.}(2015)\citenamefont
		{Banerjee}, \citenamefont {Alok}, \citenamefont {Srikanth},\ and\
		\citenamefont {Hiesmayr}}]{Banerjee}%
	\BibitemOpen
	\bibfield  {author} {\bibinfo {author} {\bibfnamefont {S.}~\bibnamefont
			{Banerjee}}, \bibinfo {author} {\bibfnamefont {A.~K.}\ \bibnamefont {Alok}},
		\bibinfo {author} {\bibfnamefont {R.}~\bibnamefont {Srikanth}}, \ and\
		\bibinfo {author} {\bibfnamefont {B.~C.}\ \bibnamefont {Hiesmayr}},\
	}\href@noop {} {\bibfield  {journal} {\bibinfo  {journal} {The European
				Physical Journal C}\ }\textbf {\bibinfo {volume} {75}},\ \bibinfo {pages}
		{487} (\bibinfo {year} {2015})}\BibitemShut {NoStop}%
	\bibitem [{\citenamefont {Cervera-Lierta}\ \emph {et~al.}(2017)\citenamefont
		{Cervera-Lierta}, \citenamefont {Latorre}, \citenamefont {Rojo},\ and\
		\citenamefont {Rottoli}}]{Cervera}%
	\BibitemOpen
	\bibfield  {author} {\bibinfo {author} {\bibfnamefont {A.}~\bibnamefont
			{Cervera-Lierta}}, \bibinfo {author} {\bibfnamefont {J.}~\bibnamefont
			{Latorre}}, \bibinfo {author} {\bibfnamefont {J.}~\bibnamefont {Rojo}}, \
		and\ \bibinfo {author} {\bibfnamefont {L.}~\bibnamefont {Rottoli}},\
	}\href@noop {} {\bibfield  {journal} {\bibinfo  {journal} {SciPost Physics}\
		}\textbf {\bibinfo {volume} {3}},\ \bibinfo {pages} {036} (\bibinfo {year}
		{2017})}\BibitemShut {NoStop}%
	\bibitem [{\citenamefont {Kerbikov}(2018)}]{Kerbikov}%
	\BibitemOpen
	\bibfield  {author} {\bibinfo {author} {\bibfnamefont {B.}~\bibnamefont
			{Kerbikov}},\ }\href@noop {} {\bibfield  {journal} {\bibinfo  {journal}
			{Nuclear Physics A}\ }\textbf {\bibinfo {volume} {975}},\ \bibinfo {pages}
		{59} (\bibinfo {year} {2018})}\BibitemShut {NoStop}%
	\bibitem [{\citenamefont {Fu}\ and\ \citenamefont {Chen}(2017)}]{Fu}%
	\BibitemOpen
	\bibfield  {author} {\bibinfo {author} {\bibfnamefont {Q.}~\bibnamefont
			{Fu}}\ and\ \bibinfo {author} {\bibfnamefont {X.}~\bibnamefont {Chen}},\
	}\href@noop {} {\bibfield  {journal} {\bibinfo  {journal} {The European
				Physical Journal C}\ }\textbf {\bibinfo {volume} {77}},\ \bibinfo {pages}
		{775} (\bibinfo {year} {2017})}\BibitemShut {NoStop}%
	\bibitem [{\citenamefont {Richter-Laskowska}\ \emph {et~al.}(2018)\citenamefont
		{Richter-Laskowska}, \citenamefont {{\L}obejko},\ and\ \citenamefont
		{Dajka}}]{Richter}%
	\BibitemOpen
	\bibfield  {author} {\bibinfo {author} {\bibfnamefont {M.}~\bibnamefont
			{Richter-Laskowska}}, \bibinfo {author} {\bibfnamefont {M.}~\bibnamefont
			{{\L}obejko}}, \ and\ \bibinfo {author} {\bibfnamefont {J.}~\bibnamefont
			{Dajka}},\ }\href@noop {} {\bibfield  {journal} {\bibinfo  {journal} {New
				Journal of Physics}\ }\textbf {\bibinfo {volume} {20}},\ \bibinfo {pages}
		{063040} (\bibinfo {year} {2018})}\BibitemShut {NoStop}%
	\bibitem [{\citenamefont {Dixit}\ \emph
		{et~al.}(2018{\natexlab{b}})\citenamefont {Dixit}, \citenamefont {Naikoo},
		\citenamefont {Banerjee},\ and\ \citenamefont {Alok}}]{Dixit}%
	\BibitemOpen
	\bibfield  {author} {\bibinfo {author} {\bibfnamefont {K.}~\bibnamefont
			{Dixit}}, \bibinfo {author} {\bibfnamefont {J.}~\bibnamefont {Naikoo}},
		\bibinfo {author} {\bibfnamefont {S.}~\bibnamefont {Banerjee}}, \ and\
		\bibinfo {author} {\bibfnamefont {A.~K.}\ \bibnamefont {Alok}},\ }\href@noop
	{} {\bibfield  {journal} {\bibinfo  {journal} {The European Physical Journal
				C}\ }\textbf {\bibinfo {volume} {78}},\ \bibinfo {pages} {914} (\bibinfo
		{year} {2018}{\natexlab{b}})}\BibitemShut {NoStop}%
	\bibitem [{\citenamefont {Nikitin}\ \emph {et~al.}(2015)\citenamefont
		{Nikitin}, \citenamefont {Sotnikov},\ and\ \citenamefont {Toms}}]{Nikitin}%
	\BibitemOpen
	\bibfield  {author} {\bibinfo {author} {\bibfnamefont {N.}~\bibnamefont
			{Nikitin}}, \bibinfo {author} {\bibfnamefont {V.}~\bibnamefont {Sotnikov}}, \
		and\ \bibinfo {author} {\bibfnamefont {K.}~\bibnamefont {Toms}},\ }\href
	{\doibase 10.1103/PhysRevD.92.016008} {\bibfield  {journal} {\bibinfo
			{journal} {Physical Review D}\ }\textbf {\bibinfo {volume} {92}},\ \bibinfo
		{pages} {016008} (\bibinfo {year} {2015})}\BibitemShut {NoStop}%
	\bibitem [{\citenamefont {Naikoo}\ \emph {et~al.}(2017)\citenamefont {Naikoo},
		\citenamefont {Alok}, \citenamefont {Banerjee}, \citenamefont {Sankar},
		\citenamefont {Guarnieri},\ and\ \citenamefont {Hiesmayr}}]{Naikoo:2017fos}%
	\BibitemOpen
	\bibfield  {author} {\bibinfo {author} {\bibfnamefont {J.}~\bibnamefont
			{Naikoo}}, \bibinfo {author} {\bibfnamefont {A.~K.}\ \bibnamefont {Alok}},
		\bibinfo {author} {\bibfnamefont {S.}~\bibnamefont {Banerjee}}, \bibinfo
		{author} {\bibfnamefont {S.~U.}\ \bibnamefont {Sankar}}, \bibinfo {author}
		{\bibfnamefont {G.}~\bibnamefont {Guarnieri}}, \ and\ \bibinfo {author}
		{\bibfnamefont {B.~C.}\ \bibnamefont {Hiesmayr}},\ }\href@noop {} {\
		(\bibinfo {year} {2017})},\ \Eprint {http://arxiv.org/abs/1710.05562}
	{arXiv:1710.05562 [hep-ph]} \BibitemShut {NoStop}%
	\bibitem [{\citenamefont {Blasone}\ \emph {et~al.}(2009)\citenamefont
		{Blasone}, \citenamefont {Dell'Anno}, \citenamefont {De~Siena},\ and\
		\citenamefont {Illuminati}}]{blasone}%
	\BibitemOpen
	\bibfield  {author} {\bibinfo {author} {\bibfnamefont {M.}~\bibnamefont
			{Blasone}}, \bibinfo {author} {\bibfnamefont {F.}~\bibnamefont {Dell'Anno}},
		\bibinfo {author} {\bibfnamefont {S.}~\bibnamefont {De~Siena}}, \ and\
		\bibinfo {author} {\bibfnamefont {F.}~\bibnamefont {Illuminati}},\
	}\href@noop {} {\bibfield  {journal} {\bibinfo  {journal} {EPL (Europhysics
				Letters)}\ }\textbf {\bibinfo {volume} {85}},\ \bibinfo {pages} {50002}
		(\bibinfo {year} {2009})}\BibitemShut {NoStop}%
	\bibitem [{\citenamefont {Birrell}\ and\ \citenamefont
		{Davies}(1984)}]{birrell}%
	\BibitemOpen
	\bibfield  {author} {\bibinfo {author} {\bibfnamefont {N.~D.}\ \bibnamefont
			{Birrell}}\ and\ \bibinfo {author} {\bibfnamefont {P.~C.~W.}\ \bibnamefont
			{Davies}},\ }\href@noop {} {\emph {\bibinfo {title} {Quantum Fields in Curved
				Space}}}\ (\bibinfo  {publisher} {Cambridge Univ. Press},\ \bibinfo {address}
	{Cambridge, UK},\ \bibinfo {year} {1984})\BibitemShut {NoStop}%
	\bibitem [{\citenamefont {Kaku}(1993)}]{kaku}%
	\BibitemOpen
	\bibfield  {author} {\bibinfo {author} {\bibfnamefont {M.}~\bibnamefont
			{Kaku}},\ }\href@noop {} {\emph {\bibinfo {title} {Quantum field theory: a
				modern introduction}}}\ (\bibinfo  {publisher} {Oxford University Press, Oxford},\
	\bibinfo {year} {1993})\BibitemShut {NoStop}%
	\bibitem [{\citenamefont {Schwinger}(2018)}]{schw}%
	\BibitemOpen
	\bibfield  {author} {\bibinfo {author} {\bibfnamefont {J.}~\bibnamefont
			{Schwinger}},\ }\href {https://books.google.co.in/books?id=HGhQDwAAQBAJ}
	{\emph {\bibinfo {title} {Particles, Sources, And Fields}}},\  (\bibinfo  {publisher} {AddisonWesley, Reading, MA},\ \bibinfo {year}
	{1973})\BibitemShut {NoStop}%
	\bibitem [{\citenamefont {Huang}\ and\ \citenamefont
		{Parker}(2009)}]{parker2009}%
	\BibitemOpen
	\bibfield  {author} {\bibinfo {author} {\bibfnamefont {X.}~\bibnamefont
			{Huang}}\ and\ \bibinfo {author} {\bibfnamefont {L.}~\bibnamefont {Parker}},\
	}\href {\doibase 10.1103/PhysRevD.79.024020} {\bibfield  {journal} {\bibinfo
			{journal} {Phys. Rev. D}\ }\textbf {\bibinfo {volume} {79}},\ \bibinfo
		{pages} {024020} (\bibinfo {year} {2009})}\BibitemShut {NoStop}%
	\bibitem [{\citenamefont {Obukhov}\ \emph {et~al.}(2009)\citenamefont
		{Obukhov}, \citenamefont {Silenko},\ and\ \citenamefont
		{Teryaev}}]{0907.4367}%
	\BibitemOpen
	\bibfield  {author} {\bibinfo {author} {\bibfnamefont {Y.~N.}\ \bibnamefont
			{Obukhov}}, \bibinfo {author} {\bibfnamefont {A.~J.}\ \bibnamefont
			{Silenko}}, \ and\ \bibinfo {author} {\bibfnamefont {O.~V.}\ \bibnamefont
			{Teryaev}},\ }\href {\doibase 10.1103/PhysRevD.80.064044} {\bibfield
		{journal} {\bibinfo  {journal} {Phys. Rev. D}\ }\textbf {\bibinfo {volume}
			{80}},\ \bibinfo {pages} {064044} (\bibinfo {year} {2009})}\BibitemShut
	{NoStop}%
	\bibitem [{\citenamefont {{Dvornikov}}(2006)}]{0601095}%
	\BibitemOpen
	\bibfield  {author} {\bibinfo {author} {\bibfnamefont {M.}~\bibnamefont
			{{Dvornikov}}},\ }\href {\doibase 10.1142/S021827180600870X} {\bibfield
		{journal} {\bibinfo  {journal} {International Journal of Modern Physics D}\
		}\textbf {\bibinfo {volume} {15}},\ \bibinfo {pages} {1017} (\bibinfo {year}
		{2006})},\ \Eprint {http://arxiv.org/abs/hep-ph/0601095}
	{arXiv:hep-ph/0601095 [hep-ph]} \BibitemShut {NoStop}%
	\bibitem [{\citenamefont {Barenboim}\ \emph {et~al.}(2002)\citenamefont
		{Barenboim}, \citenamefont {Beacom}, \citenamefont {Borissov},\ and\
		\citenamefont {Kayser}}]{baren}%
	\BibitemOpen
	\bibfield  {author} {\bibinfo {author} {\bibfnamefont {G.}~\bibnamefont
			{Barenboim}}, \bibinfo {author} {\bibfnamefont {J.~F.}\ \bibnamefont
			{Beacom}}, \bibinfo {author} {\bibfnamefont {L.}~\bibnamefont {Borissov}}, \
		and\ \bibinfo {author} {\bibfnamefont {B.}~\bibnamefont {Kayser}},\
	}\href@noop {} {\bibfield  {journal} {\bibinfo  {journal} {Physics Letters
				B}\ }\textbf {\bibinfo {volume} {537}},\ \bibinfo {pages} {227} (\bibinfo
		{year} {2002})}\BibitemShut {NoStop}%
	\bibitem [{\citenamefont {Blasone}\ \emph {et~al.}(2014)\citenamefont
		{Blasone}, \citenamefont {Dell'Anno}, \citenamefont {De~Siena},\ and\
		\citenamefont {Illuminati}}]{blasone2014}%
	\BibitemOpen
	\bibfield  {author} {\bibinfo {author} {\bibfnamefont {M.}~\bibnamefont
			{Blasone}}, \bibinfo {author} {\bibfnamefont {F.}~\bibnamefont {Dell'Anno}},
		\bibinfo {author} {\bibfnamefont {S.}~\bibnamefont {De~Siena}}, \ and\
		\bibinfo {author} {\bibfnamefont {F.}~\bibnamefont {Illuminati}},\
	}\href@noop {} {\bibfield  {journal} {\bibinfo  {journal} {EPL (Europhysics
				Letters)}\ }\textbf {\bibinfo {volume} {106}},\ \bibinfo {pages} {30002}
		(\bibinfo {year} {2014})}\BibitemShut {NoStop}%
	\bibitem [{\citenamefont {Mermin}(1990)}]{Mermin}%
	\BibitemOpen
	\bibfield  {author} {\bibinfo {author} {\bibfnamefont {N.~D.}\ \bibnamefont
			{Mermin}},\ }\href@noop {} {\bibfield  {journal} {\bibinfo  {journal}
			{Physical Review Letters}\ }\textbf {\bibinfo {volume} {65}},\ \bibinfo
		{pages} {1838} (\bibinfo {year} {1990})}\BibitemShut {NoStop}%
	\bibitem [{\citenamefont {Collins}\ \emph {et~al.}(2002)\citenamefont
		{Collins}, \citenamefont {Gisin}, \citenamefont {Popescu}, \citenamefont
		{Roberts},\ and\ \citenamefont {Scarani}}]{Collins}%
	\BibitemOpen
	\bibfield  {author} {\bibinfo {author} {\bibfnamefont {D.}~\bibnamefont
			{Collins}}, \bibinfo {author} {\bibfnamefont {N.}~\bibnamefont {Gisin}},
		\bibinfo {author} {\bibfnamefont {S.}~\bibnamefont {Popescu}}, \bibinfo
		{author} {\bibfnamefont {D.}~\bibnamefont {Roberts}}, \ and\ \bibinfo
		{author} {\bibfnamefont {V.}~\bibnamefont {Scarani}},\ }\href@noop {}
	{\bibfield  {journal} {\bibinfo  {journal} {Physical Review Letters}\
		}\textbf {\bibinfo {volume} {88}},\ \bibinfo {pages} {170405} (\bibinfo
		{year} {2002})}\BibitemShut {NoStop}%
	\bibitem [{\citenamefont {Scarani}\ and\ \citenamefont {Gisin}(2001)}]{Gisin}%
	\BibitemOpen
	\bibfield  {author} {\bibinfo {author} {\bibfnamefont {V.}~\bibnamefont
			{Scarani}}\ and\ \bibinfo {author} {\bibfnamefont {N.}~\bibnamefont
			{Gisin}},\ }\href@noop {} {\bibfield  {journal} {\bibinfo  {journal}
			{Physical Review Letters}\ }\textbf {\bibinfo {volume} {87}},\ \bibinfo
		{pages} {117901} (\bibinfo {year} {2001})}\BibitemShut {NoStop}%
	\bibitem [{\citenamefont {Svetlichny}(1987)}]{Svetlichny}%
	\BibitemOpen
	\bibfield  {author} {\bibinfo {author} {\bibfnamefont {G.}~\bibnamefont
			{Svetlichny}},\ }\href@noop {} {\bibfield  {journal} {\bibinfo  {journal}
			{Physical Review D}\ }\textbf {\bibinfo {volume} {35}},\ \bibinfo {pages}
		{3066} (\bibinfo {year} {1987})}\BibitemShut {NoStop}%
	\bibitem [{\citenamefont {Alsina}\ and\ \citenamefont
		{Latorre}(2016)}]{Alsina}%
	\BibitemOpen
	\bibfield  {author} {\bibinfo {author} {\bibfnamefont {D.}~\bibnamefont
			{Alsina}}\ and\ \bibinfo {author} {\bibfnamefont {J.~I.}\ \bibnamefont
			{Latorre}},\ }\href {\doibase 10.1103/PhysRevA.94.012314} {\bibfield
		{journal} {\bibinfo  {journal} {Physical Review A}\ }\textbf {\bibinfo
			{volume} {94}},\ \bibinfo {pages} {012314} (\bibinfo {year}
		{2016})}\BibitemShut {NoStop}%
	\bibitem [{\citenamefont {Bancal}\ \emph {et~al.}(2011)\citenamefont {Bancal},
		\citenamefont {Brunner}, \citenamefont {Gisin},\ and\ \citenamefont
		{Liang}}]{Bancal}%
	\BibitemOpen
	\bibfield  {author} {\bibinfo {author} {\bibfnamefont {J.-D.}\ \bibnamefont
			{Bancal}}, \bibinfo {author} {\bibfnamefont {N.}~\bibnamefont {Brunner}},
		\bibinfo {author} {\bibfnamefont {N.}~\bibnamefont {Gisin}}, \ and\ \bibinfo
		{author} {\bibfnamefont {Y.-C.}\ \bibnamefont {Liang}},\ }\href@noop {}
	{\bibfield  {journal} {\bibinfo  {journal} {Physical Review Letters}\
		}\textbf {\bibinfo {volume} {106}},\ \bibinfo {pages} {020405} (\bibinfo
		{year} {2011})}\BibitemShut {NoStop}%
	\end{thebibliography}

%
\end{document}